\documentclass[aps,twocolumn,groupedaddress,nofootinbib,showkeys,showpacs,amsfonts,eqsecnum]{revtex4}

\usepackage{bm}
\usepackage{amsmath}

\newcommand{\g}{{\textrm {g}}}

\newcommand{\bfa}{{\bm{a}}}
\newcommand{\bfn}{{\bm{n}}}
\newcommand{\bfm}{{\bm{m}}}
\newcommand{\bfk}{{\bm{k}}}
\newcommand{\bfx}{{\bm{x}}}
\newcommand{\bfy}{{\bm{y}}}
\newcommand{\bfh}{{\bm{h}}}
\newcommand{\bfzero}{{\bm{0}}}
\newcommand{\bfnabla}{{\bm{\nabla}}}

\newcommand{\bfF}{{\bm{F}}}
\newcommand{\bfH}{{\bm{H}}}
\newcommand{\bfT}{{\bm{T}}}

\newcommand{\bfz}{\bm{z}}

\begin{document}

\title{Existence of positive representations for complex weights}

\author{L.L. Salcedo}
%\email{salcedo@ugr.es}

\affiliation{ Departamento de F{\'\i}sica At\'omica, Molecular y
Nuclear, Universidad de Granada, E-18071 Granada, Spain }

\date{\today}

\begin{abstract}
The  necessity  of  computing  integrals  with  complex  weights  over
manifolds  with a  large number  of  dimensions, e.g.,  in some  field
theoretical  settings, poses  a problem  for  the use  of Monte  Carlo
techniques.  Here  it  is  shown  that  very  general  complex  weight
functions $P(x)$  on ${\mathbb{R}}^d$ can  be represented by  real and
positive weights $p(z)$ on ${\mathbb{C}}^d$, in the sense that for any
observable  $f$,  $\langle  f(x)  \rangle_P=\langle  f(z)  \rangle_p$,
$f(z)$ being the analytical  extension of $f(x)$.  The construction is
extended to arbitrary compact Lie groups.
\end{abstract}



\maketitle

\section{Introduction}

The computation of expectation  values in statistical mechanics and in
quantum field  theory in its functional  integral formulation requires
taking averages of functions with  a large number of variables. In the
continuous  case  this means  integration  over  manifolds with  large
dimensions.  In  such cases standard  numerical integration techniques
are no longer efficient and one  has to resort to Monte Carlo methods.
Unfortunately, in  some applications of great  practical interest such
as  lattice  quantum chromodynamics  in  the  presence  of a  baryonic
chemical potential \cite{Muroya:2003qs,Lombardo:2004uy}, the Boltzmann
weight to be used in the  averages is not positive or even real.  This
fact prevents a straightforward  application of the Monte Carlo method
in these cases.

For a  (real and positive) probability density  function (PDF) $P(x)$,
expectation values  can be estimated by  a pure Monte  Carlo method in
which  the   points  are  independently  sampled   from  $P(x)$.   The
dispersion in the estimate of  $\langle f \rangle_P$ in the pure Monte
Carlo  method is $\sigma_P(f)/\sqrt{N}$  where $\sigma^2_P(f)$  is the
variance of $f$  and $N$ is the number  of independent sampling points
\cite{Negele:1988vy,Madras:2002bk}. If sampling  $P(x)$ is very costly
it may be more convenient to  use the so called reweighting method, in
which an auxiliary PDF $P_0(x)$  is sampled instead, making use of the
identity
\begin{eqnarray}
\langle f\rangle_P &=&
\frac{\langle fPP_0^{-1}\rangle_{P_0}}
{\langle PP_0^{-1}\rangle_{P_0}}
\,.
\label{eq:1.1}
\end{eqnarray}
For a  generic observable $f(x)$ this  method is less  efficient as it
suffers from the importance sampling  problem: less points fall in the
relevant region (i.e., where  $P(x)$, rather than $P_0(x)$, is large),
and  the dispersion $\sigma_P(f)/\sqrt{N_{\text{eff}}}$  increases.  A
typical signal of  this problem is the presence  of large fluctuations
in  both the  numerator  and the  denominator  in (\ref{eq:1.1})  when
$P_0(x)$ and $P(x)$ are too different.
 
Nevertheless, the reweighting method  can be translated immediately to
the cases in which the  weight function $P(x)$ is complex.  (With some
abuse of language such complex density functions are still referred to
as  PDFs.)    A  positive  $P_0(x)$  is  chosen   (a  standard  choice
being\footnote{Taking  $P_0(x)$ to be  proportional to  $|P(x)|$ needs
not be the optimal choice in practice, see e.g.  \cite{Fodor:2001pe}.}
$P_0(x)= \lambda |P(x)|$) and  used to generate the samples.  Although
the importance  sampling problem is  not well defined here  in general
(being  $P(x)$ complex, sampling  it is  meaningless), the  same large
fluctuation problem which  appeared in the case of  positive $P(x)$ is
present  here as a  sign (or  rather phase)  problem, since  often the
average of the phase,  $\langle P/|P|\rangle_{|P|}$, is very small but
not  its  variance.\footnote{See  \cite{Muroya:2003qs,Lombardo:2004uy}
for alternative  techniques, such as Taylor expansion,  in the context
of   lattice  QCD   and  \cite{Anagnostopoulos:2001yb,Azcoiti:2002vk,%
Ambjorn:2002pz,Moreira:2003,Markham:1996pu,Baaquie:2000rq}  for  other
approaches to the complex action problem.}

There are cases, however, where a  sampling of a complex $P(x)$ can be
given  a meaning.   A  clear instance  of  this is  a  PDF defined  on
$\mathbb{R}$  such that $P(x)=P_0(x-ia)$,  where $P_0(x)$  is positive
(for real  $x$) and analytic  on a region containing  $\mathbb{R}$ and
$\mathbb{R}-ia$.   Then for  $f(z)$  analytic on  a region  containing
$\mathbb{R}$   and  $\mathbb{R}+ia$,   $\langle  f(x)\rangle_P=\langle
f(x+ia)\rangle_{P_0}$.   In this case  the Monte  Carlo method  can be
applied using  as sampling points  $z_k=x_k+ia$, $k=1,\ldots,N$, where
the $x_k$ are generated from $P_0(x)$, and taking as observable $f(z)$
(the analytic extension of $f(x)$).  This example suggests a technique
consisting of sampling the complex plane, or more generally a suitable
complexified version of the original manifold where the complex $P(x)$
is defined, using  an appropriate real and positive  $p(z)$, and trade
the  computation  of  $\langle  f(x)\rangle_P$  by  that  of  $\langle
f(z)\rangle_p$. In practice this program  has been applied by means of
the        so       called        complex        Langevin       method
\cite{Parisi:1984cs,Klauder:1984}.  For  positive $P(x)$ this approach
produces a  random walk  which asymptotically samples  the probability
density function.   This property is easily shown  from the associated
Fokker-Planck  equation  of  which  $P(x)$ is  the  stable  stationary
solution.  The stochastic differential  equation can be applied to the
complex case  using $P(z)$, the  analytical extension of  $P(x)$.  For
some $P(x)$ it  can be shown that the random walk  will reach a stable
equilibrium  which   samples  a   certain  $p(z)$  with   the  correct
expectation values \cite{Haymaker:1989hn}.   In those cases the method
is very useful  since then essentially the same  local algorithms used
in  the real  case apply.   Unfortunately, for  many  relevant complex
$P(x)$  there is  no steady  solution  or, if  there is,  it does  not
display   the   correct   expectation  values   \cite{Ambjorn:1985iw,%
Ambjorn:1986fz,Ambjorn:1986mf,Hamber:1985qh,Flower:1986hv,%
Schoenmaker:1987fk,Bilic:1987fn,Soderberg:1987pd,Haymaker:1987ey,%
Okamoto:1988ru,Okano:1991tz,Gausterer:1992jz,Salcedo:1993tj,%
Fujimura:1993cq,Gausterer:1998jw,Berges:2005yt,Berges:2006xc}.      As
shown in \cite{Salcedo:1993tj} the latter problem arises from the fact
that the Fokker-Planck equation admits  more solutions in the space of
distributions than  in the space of ordinary  functions.  However, the
complex Langevin method guarantees at  most that the PDF obtained from
projection of $p(z)$ on the  real axis is a distributional solution of
the  Fokker-Planck equation,  not  that it  should  coincide with  the
original  $P(x)$.  In  essence the  problem is  that in  that approach
everything  depends  on  $P(z)$  and  this  function  does  not  favor
integration  along the  real axis  from integration  along  many other
curves on the complex plane (which  can be viewed as a real axis after
an  analytic  change  of  variables).  Regrettably,  unlike  the  real
Langevin  case, no  practical criterion  is known  to decide  a priori
whether a given $P(x)$ will  produce a steady random walk (rather than
a state  that looks stationary  on a finite time  computer simulation)
and if so,  whether such steady solution is  really sampling the input
PDF $P(x)$. (See however \cite{Lee:1994sw,Adami:2000fs}.)

Given the  limited success of  the complex Langevin method,  a natural
question is to  what extent a given complex  weight can be represented
at all by  means of any suitable ordinary (i.e.,  positive) PDF on the
complexified    manifold.    A   partial    answer   was    given   in
\cite{Weingarten:2002xs}  where  it  was  shown that  one  dimensional
complex PDFs  admit such representation.   In \cite{Salcedo:1996sa} it
was  shown that  all complex  weights  on $\mathbb{R}^d$  of the  type
Gaussian  times  polynomial  of  any  degree  and  in  any  number  of
dimensions    are   also   representable    by   positive    PDFs   on
$\mathbb{C}^d$. Since  this set  is dense in  $L^2(\mathbb{R}^d)$ this
suggests that  representability is a  quite general property.   In the
present work  we extend those results by  proving representability for
smooth  complex  PDFs  on  $\mathbb{R}^d$, which  are  either  rapidly
decreasing  at  infinity  or   periodic.   The  latter  case  is  then
generalized to  smooth PDFs  defined on any  compact matrix  Lie group
manifold.

\section{Representation for PDFs of the type charge-dipole pair}

Let $P(\bfx)$ be the complex ``probability'' density function for
which we want to compute expectation values of ``observables'' $f(\bfx)$,
\begin{equation}
\langle f\rangle_P := 
\frac{\int  d\mu(\bfx) P(\bfx) f(\bfx)}{\int d\mu(\bfx) P(\bfx)}
\,.
\label{eq:2.1}
\end{equation}
$P$ and  $f$ are complex in  general, $d\mu(\bfx)$ is  a positive measure
and the  normalization $\int d\mu(\bfx)  P(\bfx)$ is different from  zero. A
positive representation  of $P(\bfx)$  is a real  nonnegative probability
density  function  $p(\bfz)$ defined  on  the  complex  extension of  the
original real manifold, with  some positive measure $d\mu(\bfz)$, in such
a way that
\begin{equation}
\langle f\rangle_P = \langle f\rangle_p := \frac{\int d\mu(\bfz) \,p(\bfz)
f(\bfz)}{\int d\mu(\bfz) \, p(\bfz)} \,,
\label{eq:2.2}
\end{equation}
where $f(\bfz)$ stands for the analytical extension of $f(\bfx)$. Of
course a precise definition requires to specify the space of allowed
test functions $f$.  For a real manifold such as ${\mathbb{R}}^d$ a
usual choice is that of the set of polynomials. For complex
probabilities which are periodic (real manifold equivalent to a torus)
the natural test functions are of the type $e^{i\bfn\cdot\bfx}$.  In
general, the larger the set of test functions the smaller the number
of representable complex probabilities. Hereafter we will assume that
$P(\bfx)$ and $p(\bfz)$ are normalized, i.e., the denominators in
(\ref{eq:2.1}) and (\ref{eq:2.2}) are unity.

The key  idea of the construction is  to decompose $P(\bfx)$ as  a sum of
simpler PDFs
\begin{equation}
P(\bfx)= \sum_n P_n(\bfx)
\end{equation}
all of them with positive normalization, 
or, equivalently,
\begin{equation}
P(\bfx)= \sum_n w_n P_n(\bfx)
\end{equation}
with normalized $P_n$ and $w_n\ge0$, and then find positive and
normalized representations $p_n(\bfz)$ for the $P_n(\bfx)$ so that
\begin{equation}
p(\bfz)= \sum_n w_n p_n(\bfz)
\end{equation}
provides the desired positive representation of $P(\bfx)$.

A  suitable choice  is  to take  the  $P_n$ of  the  form Dirac  delta
distribution plus a derivative of a Dirac delta. If $P$ is viewed as a
hypothetical (complex) charge distribution with positive total charge,
it can  be decomposed as  a sum of  simpler distributions of  the form
positive  charge (delta  distribution)  plus a  dipole (derivative  of
delta)  at   the  same  point,   with  complex  dipolar   moment.   In
${\mathbb{R}}^d$ this is
\begin{equation}
Q(\bfx;\bfh)=\delta(\bfx)+\bfh \cdot \bfnabla\delta(\bfx) \,,\quad
\bfx\in{\mathbb{R}}^d\,,\quad \bfh\in{\mathbb{C}}^d \,.
\label{eq:2.3}
\end{equation}
As  shown constructively  in  \cite{Salcedo:1996sa} such  distribution
admits  a   positive  representation   of  the  form   Gaussian  times
polynomial. A similar but simpler solution is considered here.

Let us start by considering the one-dimensional PDF
\begin{equation}
Q_1(x) = \delta(x)+\delta^\prime(x),\quad x\in {\mathbb{R}}.
\label{eq:2.5}
\end{equation}
It  can  be represented  on  the complex  plane  by  the positive  and
normalized PDF
\begin{equation}
q_1(z)  = \frac{1}{8\pi}\left|1-\frac{z}{2}\right|^2e^{-|z/2|^2},\quad
z\in {\mathbb{C}},
\end{equation}
that is
\begin{eqnarray}
\int_{\mathbb{C}} d^2z  \,  q_1(z)  \,  z^n &=& 
\int_{\mathbb{R}}  dx  \,  Q_1(x)  \,  x^n  
\nonumber \\
&=& \delta_{n,0} - \delta_{n,1}\,,
\quad
n=0,1,2,\ldots
\end{eqnarray}
where $d^2z=d{\text{Re}\,z}\,d{\text{Im}\,z}$  is the Lebesgue measure
on  $\mathbb{R}^2$. This can  be checked  by direct  computation using
polar coordinates, or alternatively, writing $q_1$ as
\begin{eqnarray}
q_1(z) &=&  \left(1+\frac{\partial}{\partial z}
+\frac{\partial}{\partial z^*}
+2\frac{\partial}{\partial z}\frac{\partial}{\partial z^*}\right)g(z)
\,,
\nonumber \\
g(z) &:=& \frac{1}{4\pi}e^{-|z/2|^2} \,,
\end{eqnarray}
integrating by parts and noting that $\langle z^n\rangle_g=
\delta_{n,0}$.

Now a positive representation of $Q(\bfx;\bfh)$ is easily obtained as
follows:
\begin{equation}
q(\bfz;\bfh)=  
 \int_{\mathbb{C}} d^2 z_1 \, q_1(z_1)\, \delta(\bfz -  z_1 \bfh) \,.
\label{eq:2.8}
\end{equation}
($\delta(\bfz-\bfz_0)$ being the  Dirac delta distribution at $\bfz_0$
of the measure $d^{2d}z$.) Indeed,
\begin{eqnarray}
\langle f(\bfz) \rangle_{q_\bfh} &=&
\int_{\mathbb{C}^d} d^{2d}z \, f(\bfz) \, q(\bfz;\bfh)
\nonumber \\
&=&
\int_{\mathbb{C}^d} d^{2d}z \, f(\bfz)  \int_{\mathbb{C}} d^2 z_1 \, 
q_1(z_1)\, \delta(\bfz -  z_1 \bfh)
\nonumber \\
&=&
\int_{\mathbb{C}} d^2 z_1  f(z_1 \bfh)   \, q_1(z_1)
\nonumber \\
&=&
\int_{\mathbb{R}} d x_1  f(x_1 \bfh)   \,  (\delta(x_1)+\delta^\prime(x_1))
\nonumber \\
&=&
f(\bfzero) -\bfh\cdot\bfnabla f (\bfzero)
\nonumber \\ 
&=&
\int_{\mathbb{R}^d} d^dx \, f(\bfx) \, Q(\bfx;\bfh)
\nonumber \\ 
&=& \langle f(\bfx) \rangle_{Q_\bfh} \,.
\end{eqnarray}
In  the fourth  equality  we  have used  that  $f(z_1  \bfh)$  depends
analytically on  $z_1$ and $q_1(z)$  is a representation  of $Q_1(x)$.

The support of the distribution $q(\bfz;\bfh)$ is at most a plane, and
Gaussian-like on that plane.  From the Monte Carlo point of view this
is considerably more efficient than the solutions presented in
\cite{Salcedo:1996sa} which had a larger size (Gaussian-like with non
vanishing width in all $2d$ directions).

\section{Periodic PDFs}
\label{sec:3}

Next let us apply this  technique to find positive representations for
periodic  PDFs functions  on  ${\mathbb{R}}^d$. The  advantage of  the
periodic   case  is   that  the   manifold  is   effectively  compact,
topologically  a $d$-dimensional torus.   Specifically we  take $P(\bfx)$
with period $1$ in each direction,
\begin{equation}
P(\bfx+\bfn) = P(\bfx) \,, \quad \bfx\in  {\mathbb{R}}^d\,, \quad 
\bfn\in  {\mathbb{Z}}^d\,,
\end{equation}
and  normalized  to  unity   with  measure  $d\mu(\bfx)=d^dx$  on  the
$d$-dimensional  torus   $T^d=[0,1]^d$  (suitably  compactified).   In
addition,  we  assume $P(\bfx)$  to  be  a  smooth (i.e.   $C^\infty$)
function on the torus.\footnote{Smoothness  is invoked to simplify the
treatment. Likely the constructions  presented here can be extended to
suitable  spaces  of  distributions.}   A positive  representation  of
$P(\bfx)$   is  a   smooth   nonnegative  function   $p(\bfz)$  on   $
{\mathbb{C}}^d$ which is periodic (in the real directions)
\begin{equation}
p(\bfz+\bfn) = p(\bfz) \,, \quad \bfz\in  {\mathbb{C}}^d\,, \quad 
\bfn\in  {\mathbb{Z}}^d\,,
\end{equation}
and such that $\langle e^{2\pi i \bfn\cdot\bfx}\rangle_P=
\langle e^{2\pi i \bfn \cdot\bfz}\rangle_p$ for all 
$\bfn\in  {\mathbb{Z}}^d$, that is
\begin{equation}
\int_{T^d} d^dx \,P(\bfx)e^{2\pi i \bfn\cdot\bfx} = 
\int_{\tilde{T}^d} d^{2d}z \, p(\bfz)e^{2\pi i \bfn\cdot\bfz}
\,,\quad
\bfn\in  {\mathbb{Z}}^d\,,
\end{equation}
where  the integration manifold  of $\bfz$  is the  complexified torus
$\tilde{T}^d=[0,1]^d\times   {\mathbb{R}}^d$.     This   manifold   is
non-compact in  the imaginary direction.  For subsequent  use, we note
that the periodic version of $Q(\bfx;\bfh)$,
\begin{equation}
Q^p(\bfx;\bfh) = \sum_{\bfn\in  {\mathbb{Z}}^d} Q(\bfx+\bfn;\bfh)
\label{eq:2.13}
\end{equation}
admits the positive representation
\begin{equation}
q^p(\bfz;\bfh) = \sum_{\bfn\in {\mathbb{Z}}^d} q(\bfz+\bfn;\bfh) \,,
\end{equation}
the  sum  being convergent  (as  a  distribution).  All the  functions
considered, being periodic, are  well defined (one-valued) on $T^d$ or
$\tilde{T}^d$.

We can decompose $P(\bfx)$ in the form
\begin{equation}
P(\bfx) = P_0(\bfx) + \bfnabla\cdot \bfF(\bfx)
\label{eq:3.6}
\end{equation}
where $P_0(\bfx)$ is chosen to be everywhere strictly positive, smooth
(on  the  torus)  and  normalized  to unity.   A  sensible  choice  is
$P_0(\bfx)=1$, however we will keep  the possibility of a more general
choice to  maintain the analogy with  the non compact  case, below. By
Hodge's  decomposition  theorem  \cite{Nakahara:1990th},  there  is  a
complex vector  field $\bfF(\bfx)$ which  is also smooth on  the torus
(smooth and  periodic on ${\mathbb{R}}^d$).  Such  $\bfF(\bfx)$ is non
unique.  A natural solution is just the electric field-like one: let
\begin{equation}
\rho(\bfx):= P(\bfx)-P_0(\bfx) \,,
\end{equation}
then (\ref{eq:3.6}) becomes
\begin{equation}
\bfnabla\cdot\bfF(\bfx)= \rho(\bfx)\,.
\end{equation}
The electric field-like solution is
\begin{equation}
\bfF(\bfx)=\bfnabla \Phi(\bfx)\,,\quad
\label{eq:3.7}
\end{equation}
where $\Phi(\bfx)$ is a smooth periodic solution of
\begin{equation}
\bfnabla^2\Phi(\bfx)=\rho(\bfx) 
\,.
\end{equation}
More explicitly,
\begin{eqnarray}
\rho(\bfx) &=& \sum_{\bfn\in {\mathbb{Z}^d,\bfn\not=\bfzero}}
\rho_\bfn e^{2\pi i \bfn\cdot\bfx} 
\,,
\nonumber \\
\Phi(\bfx) &=& \sum_{\bfn\in {\mathbb{Z}^d,\bfn\not=\bfzero}}
\frac{1}{(2\pi i)^2}\frac{\rho_\bfn}{\bfn^2} 
e^{2\pi i \bfn\cdot\bfx} 
\,,
\nonumber \\
\bfF(\bfx) &=& \sum_{\bfn\in  {\mathbb{Z}}^d,\bfn\not=\bfzero} 
\frac{\bfn}{2\pi i}\frac{\rho_\bfn }{\bfn^2}
e^{ 2\pi i \bfn\cdot\bfx}
\,.
\end{eqnarray}
Note that $\bfF(\bfx)$ is smooth because $\rho(\bfx)$ is smooth (i.e.,
$|\rho_\bfn|$ decreases for large $\bfn$ faster that any inverse
power). $\rho_\bfzero$ vanishes due to $\int d^dx\rho(\bfx)=0$, since
$P(\bfx)$ and $P_0(\bfx)$ are both normalized to unity.

Having a valid $\bfF(\bfx)$, (\ref{eq:3.6}) can be rewritten as
\begin{equation}
P(\bfx) = \int_{\mathbb{R}^d} d^dy \, P_0(\bfy) \left(\delta(\bfx-\bfy)
+\bfH(\bfy) \cdot
\bfnabla\delta(\bfx-\bfy)\right)
\end{equation}
with
\begin{equation}
\bfH(\bfy)=\frac{\bfF(\bfy)}{P_0(\bfy)} \,.
\label{eq:2.16}
\end{equation}
The vector field $\bfH(\bfx)$ is also smooth on $T^d$. In turn this can be
rewritten as
\begin{eqnarray}
P(\bfx) &=& \int_{\mathbb{R}^d} d^dy \, P_0(\bfy) Q(\bfx-\bfy;\bfH(\bfy))
\nonumber \\
 &=& \int_{T^d} d^dy \, P_0(\bfy) Q^p(\bfx-\bfy;\bfH(\bfy))
\end{eqnarray}
where $Q^p(\bfx;\bfH(\bfy))$ is just $Q^p_{\bfh}(\bfx)$ of
(\ref{eq:2.13}) with $\bfh=\bfH(\bfy)$.  Therefore, $P(\bfx)$ has been
written as a sum of charge-dipole distributions of the type
(\ref{eq:2.3}).  A positive representation of $P(\bfx)$ is thus
readily found:
\begin{eqnarray}
p(\bfz) &=& \int_{T^d} d^dy \, P_0(\bfy) \, q^p(\bfz-\bfy;\bfH(\bfy))
\nonumber \\
&=&
\int_{\mathbb{R}^d} d^dy \, P_0(\bfy) \, q(\bfz-\bfy;\bfH(\bfy))
\label{eq:2.17}
\end{eqnarray}
where $q^p(\bfz;\bfH(\bfy))$ or $q(\bfz;\bfH(\bfy))$ are just
$q^p(\bfz;\bfh)$ or $q(\bfz;\bfh)$ with $\bfh=\bfH(\bfy)$.  The
positive representation $p(\bfz)$ is also smooth and periodic on
${\mathbb{C}^d}$ and so smooth on $\tilde{T}^d$.

\section{Positive representations for PDFs on compact Lie groups}

Let us now consider the extension of the previous construction to Lie
group manifolds.  Such manifolds appear often in applications such as
lattice gauge theories.  Specifically, as real manifold we take a
connected and compact Lie group $G$ of dimension $d$, which for
convenience will be taken as a matrix group. Hence all elements are of
the form $g=\exp(\bfx\cdot \bfT)$ where $\bfx\in{\mathbb{R}^d}$ and
the $d$ matrices $\bfT$ define a basis of the $d$-dimensional Lie
algebra.  The complexified manifold is the connected but non-compact
Lie group $\tilde{G}=\{\tilde{g}=\exp(\bfz\cdot\bfT),
\bfz\in{\mathbb{C}^d}\}$.  A compact Lie group $G$ admits a two-sided
invariant metric $\g_{\mu\nu}$ which endows $G$ with a Riemannian
manifold structure \cite{Barut:1986bk}.  The metric can be normalized
so that the corresponding volume element, $d^dx\sqrt{\g}$, is the Haar
measure of $G$ normalized to unity
\cite{Choquet-Bruhat:1982bk}.\footnote{$\g=\det\g_{\mu\nu}$ and
$x^\mu$ are just any local coordinates, not necessarily the normal
coordinates appearing in $e^{\bfx\cdot\bfT}$.}  This measure is used
in the evaluation of expectation values.  For expectation values on
$\tilde{G}$ we also take its Haar measure.

A natural set  of test functions, is that  of arbitrary polynomials of
the matrix elements of  $g\in G$, $f(g)=f(\{g_{ij}\})$. The analytical
extension  of these polynomials  corresponds just  to replace  $g$ with
$\tilde{g}$, i.e., $f(\{\tilde{g}_{ij}\})$.

The construction of a positive representation $p(\tilde{g})$ of a
smooth normalized complex PDF $P(g)$ on $G$ is as follows.  Let
$P_0(g)$ be strictly positive, smooth and normalized (for instance
$P_0=1$), then the difference $P-P_0$ integrates to zero on $G$ and,
by Hodge decomposition in a compact manifold, we can write
(using differential geometry notation \cite{Nakahara:1990th})
\begin{equation}
P(g)=P_0(g)-d^\dagger F(g),
\end{equation}
where $F(g)=F_\mu(g)dx^\mu$ is a one-form on $G$. Equivalently,
\begin{equation}
P(g)=P_0(g)+\nabla_\mu F^\mu(g),
\label{eq:4.3}
\end{equation}
where $\nabla_\mu$ is the covariant derivative on $G$ as a Riemannian
manifold, with the usual Levi-Civita connection, and $F^\mu(g)
=\g^{\mu\nu}(g)F_\nu(g)$.  The smooth vector field $F^\mu(g)$ is not
unique and the electric field-like solution can be adopted, for
instance, \cite{Nakahara:1990th}:
\begin{eqnarray}
F^\mu(g)=\nabla^\mu(\nabla^2)^{-1}(P(g)-P_0(g))\,.
\end{eqnarray}

Using (\ref{eq:4.3}) and upon integration by parts,
\begin{eqnarray}
\langle f(g)\rangle_P &=&
\int d\mu(g) P(g) f(g)
\nonumber \\
&=&
\int d\mu(g) P_0(g) \big(1- H^\mu(g) \partial_\mu\big) f(g)
\,,
\label{eq:2.23}
\end{eqnarray}
with
\begin{equation}
H^\mu(g)= P_0(g)^{-1}F^\mu(g)  \,.
\end{equation}

Clearly, $H^\mu(g) \partial_\mu f(g)$ describes an infinitesimal point
transformation, and this can be implemented as an infinitesimal left
translation by means of a smooth field ${\cal H}(g)$ taking values on
the Lie algebra:
\begin{equation}
H^\mu(g) \partial_\mu f(g)=
\frac{d}{dx_1} f(e^{x_1{\cal H}(g)}g)\Big|_{x_1=0} \,.
\end{equation}
With the  help of the  distribution $Q_1(x_1)$ in  (\ref{eq:2.5}), the
expression in (\ref{eq:2.23}) can then be rewritten as
\begin{eqnarray}
\langle f(g)\rangle_P=
\int d\mu(g) P_0(g)\int_{\mathbb{R}}dx_1 Q_1(x_1)
f(e^{x_1{\cal H}(g)}g) \,.
\end{eqnarray}
Since the dependence on $x_1$ in $f$ is analytic, we can use the
positive representation of $Q_1(x_1)$ on $\mathbb{C}$, $q_1(z_1)$, to
write
\begin{eqnarray}
\langle f(g)\rangle_P=
\int d\mu(g) P_0(g)\int_{\mathbb{C}}d^2z_1 q_1(z_1) f(e^{z_1{\cal H}(g)}g) \,.
\end{eqnarray}
This  allows  to  express  the  expectation  value  using  a  positive
representation on $\tilde{G}$,
\begin{eqnarray}
\langle f(g)\rangle_P=
\langle f(\tilde{g})\rangle_p :=
\int d\mu(\tilde{g}) \,p(\tilde{g})\,f(\tilde{g}) \,,
\end{eqnarray}
with
\begin{eqnarray}
p(\tilde{g})=
\int d\mu(g) P_0(g)\,q(\tilde{g},g;{\cal H}(g))
\end{eqnarray}
and
\begin{eqnarray}
q(\tilde{g},g;{\cal H}(g))= 
\int_{\mathbb{C}}d^2z_1 q_1(z_1) 
\delta(e^{z_1{\cal H}(g)}g,\tilde{g}) \,.
\end{eqnarray}
Here   $\delta(\tilde{g}_0,\tilde{g})$   denotes   the   Dirac   delta
distribution at  $\tilde{g}_0$ for the  Haar measure $d\mu(\tilde{g})$
on $\tilde{G}$. The extension to non connected $G$ is obvious.

\section{Positive representations for PDFs on $\mathbb{R}^d$}
\label{sec:5}

For PDFs on $\mathbb{R}^d$ the previous construction based on a
superposition of charge-dipole pairs can also be carried out, but it
is technically more involved due to the lack of compactness of the
real manifold.  We will assume the normalized complex PDF $P(\bfx)$ to
be in Schwartz space, i.e.  smooth (infinitely differentiable) and
rapidly decreasing at infinity ($P(\bfx)$ and all its derivatives go
to zero faster than any inverse power of $|\bfx|$).  The space of test
functions can then be chosen as the set of polynomials of $\bfx$.

We choose a suitable positive and normalized $P_0(\bfx)$ and write
$P(\bfx)$ as
\begin{equation}
P(\bfx)= P_0(\bfx) + \bfnabla\cdot\bfF(\bfx) \,.
\end{equation}
Proceeding formally we then obtain a positive representation with
\begin{equation}
p(\bfz) = \int_{\mathbb{R}^d} d^dx \, P_0(\bfx) \, 
\int_{\mathbb{C}} d^2 z_1 \, 
q_1(z_1)\, \delta(\bfz - \bfx - z_1 \bfH(\bfx))
\,,
\label{eq:5.2}
\end{equation}
with $\bfH(\bfx)=\bfF(\bfx)/P_0(\bfx)$.

There is a number of issues to be considered in this construction,
such as the existence of a suitable $P_0(\bfx)$ and of the vector
field $\bfF(\bfx)$, the convergence of $p(\bfz)$ as defined in
(\ref{eq:5.2}), since the integral is on a non compact manifold, and
finally, the convergence of the momenta $|\bfz|^n$ of $p(\bfz)$ for
all non negative $n$.

Assuming by the moment that suitable $P_0(\bfx)$ and $\bfF(\bfx)$
exist, the two latter points will be fulfilled a fortiori provided the
set of integrals
\begin{equation}
I_\bfn = \int_{\mathbb{C}^d} d^{2d}z \, p(\bfz) \, |\bfz^\bfn|
\end{equation}
exist, where $\bfz^\bfn:=\prod_{\mu=1}^d z_\mu^{n_\mu}$ and $n_\mu$
are nonnegative integers. Equivalently, the integrals
\begin{equation}
I_\bfn =  
\int_{\mathbb{R}^d} d^dx \, P_0(\bfx) \, \int_{\mathbb{C}} d^2 z_1 \, 
q_1(z_1)\,
|(\bfx+z_1 \bfH(\bfx))^\bfn| \,
\end{equation}
should be convergent.  We will take $P_0(\bfx)$ also in Schwartz space
and momentarily assume that $\bfH(\bfx)$ is smooth. In this case the
only problem of convergence may can from the large $|\bfx|$ or large
$|z_1|$ sectors of the integral.  Since $P_0(\bfx)$ and $q_1(z_1)$ are
rapidly decreasing, {\em convergence is ensured provided $\bfH(\bfx)$
is bounded by a polynomial}. This condition turns out to be rather
restrictive since it implies that $\bfF(\bfx)$ should be rapidly
decreasing and moreover it should go to zero at a rate not much slower
than $P_0(\bfx)$ itself.

Since both $P(\bfx)$ and $P_0(\bfx)$ are smooth and rapidly
decreasing, so is their difference, $\rho$
\begin{equation}
\rho(\bfx):= P(\bfx)-P_0(\bfx) \,,
\end{equation}
and the equation on
$\bfF(\bfx)$ is
\begin{equation}
\bfnabla\cdot\bfF(\bfx)= \rho(\bfx) \,.
\end{equation}
The electric field-like solution, (\ref{eq:3.7}), exists and is smooth
but unfortunately, it will not be rapidly decreasing at infinity, in
general: Because $\rho(\bfx)$ is in Schwartz space so is its Fourier
transform, ${\tilde\rho}(\bfk)$.  Then,
$(\bfk/|\bfk|^2){\tilde\rho}(\bfk)$ is rapidly decreasing but not
necessarily smooth at $\bfk=0$, correspondingly, its Fourier transform
is smooth but not rapidly decreasing, in general.  Indeed, unless
$\rho(\bfx)$ has radial symmetry, the electric field will have the (in
three dimensions) well-known multipolar contributions which fall as an
inverse power for large $\bfx$ \cite{Jackson:1998bk}.  To find a
suitable $\bfF(\bfx)$ it is better to view this quantity as (minus) a
dipolar density so that its divergence is the charge density
$\rho(\bfx)$.  Because the total charge carried by $\rho(\bfx)$ is
zero it should be possible to build it as a superposition of dipoles
with complex dipolar moment.  Moreover, if $\rho(\bfx)$ is smooth or
rapidly decreasing or of compact support, it should be possible to
choose $\bfF(\bfx)$ with the same properties (the support of
$\bfF(\bfx)$ being larger than that of $\rho(\bfx)$ in general).  A
suitable solution is as follows
\begin{equation}
\bfF_\bfa(\bfx)= -(\bfx-\bfa)\int_1^\infty d\lambda \,\lambda^{d-1}
\rho(\lambda(\bfx-\bfa)+\bfa) \,,\quad
\bfx\not=  \bfa
\label{eq:5.5}
\end{equation}
where the  basepoint $\bfa$ is  an arbitrary point in  $\mathbb{R}^d$. 
Intuitively this solution
corresponds to an arrangement of linear chains of dipoles, all chains
starting at $\bfa$ and ending at each of the points in the support of
$\rho$, in which each ``positive'' charge of one dipole is canceled by
the ``negative'' one of the next dipole in the chain.  The total
charge at $\bfa$ equals that in $\rho(\bfx)$ which is zero.

It is immediate to verify that $\bfnabla\cdot \bfF_\bfa(\bfx)=
\rho(\bfx)$ for all $\bfx\not= \bfa$. Also it is clear that
$\bfF_\bfa(\bfx)$ is rapidly decreasing at infinity (or of compact
support if $\rho(\bfx)$ is) and smooth at all points except the
basepoint.  At $\bfx=\bfa$ the solution $\bfF_\bfa(\bfx)$ will not be
smooth in general, however this problem is easily fixed by taking a
smooth average over $\bfa$,
\begin{equation}
\bfF(\bfx)= 
-\int_1^\infty d\lambda \,\lambda^{d-1}
\!\!
\int_{\mathbb{R}^d} d^da\, C(\bfa)
(\bfx-\bfa)
\rho(\lambda(\bfx-\bfa)+\bfa),
\label{eq:5.6}
\end{equation}
where we choose the weighting function $C(\bfa)$ to be smooth, of
compact support, positive and normalized to unity. Upon taking the
average over $\bfa$, the Fourier transform ${\tilde \bfF}(\bfk)$ can
be shown to be rapidly decreasing, in addition to smooth, so
$\bfF(\bfx)$ belongs to Schwartz space.\footnote{Of course, here we
are addressing the generic case.  In many practical cases this
complicated construction is not needed.  For instance, if $\rho(\bfx)$
is of the type Gaussian times polynomial, it is immediate, by going to
Fourier space, to find a valid $\bfF(\bfx)$ which is also of the type
Gaussian times polynomial.}

The remaining issue is whether $P_0(\bfx)$ can be chosen so that
$\bfH(\bfx)=\bfF(\bfx)/P_0(\bfx)$ is polynomially bounded.  Since the
discussion is rather technical, this is shown in the Appendix. Here we
only note that $P_0(\bfx)$ should not go to zero too quickly for large
$|\bfx|$, as compared to $P(\bfx)$.  Otherwise the falloff of
$\rho(\bfx)$ and $\bfF(\bfx)$ would be dominated by that of $P(\bfx)$
and the ratio $\bfF(\bfx)/P_0(\bfx)$ could fail to be bounded by a
polynomial.

\section{Alternative constructions}

The  positive representation of  a given  complex PDF  is by  no means
unique  \cite{Salcedo:1996sa}.   For  periodic  density  functions  we
present here alternative positive  representations to those studied in
section \ref{sec:3}.  They have the virtue of being quite localized on
${\mathbb{C}^d}$ or $\tilde{T}^d$. This is convenient for their use in
Monte   Carlo  integration  since   the  localization   decreases  the
fluctuations in the average over samples.

The construction is based on decomposing the periodic and smooth
$P(\bfx)$ as a weighted sum (with positive weight) of simpler PDFs of
the form
\begin{equation}
P_\bfn(\bfx)=1+A_\bfn e^{2\pi i \bfn\cdot\bfx} \,,\quad 
\bfn\in \mathbb{Z}^d
\,,\quad \bfn\not=\bfzero\,.
\end{equation}
If the normalization of $P(\bfx)$ is positive such decomposition is
clearly always possible after a discrete Fourier decomposition.
$P_\bfn(\bfx)$ has expectation values
\begin{equation}
\langle e^{-2\pi i \bfm \cdot\bfx}\rangle_{P_\bfn} = \left\{
\begin{matrix}
1 & \text{for~}\bfm= \bfzero \hfill  \cr
A_\bfn & \text{for~}\bfm= \bfn \hfill \cr
0 & \text{otherwise} \hfill 
\end{matrix}
\right. \,,\quad \bfm\in \mathbb{Z}^d\,.
\end{equation}
Actually the parameter $A_\bfn$ is redundant.  Indeed, the expectation
values of two PDFs $p(\bfz)$ and $p^\prime(\bfz)=p(\bfz+\bfa)$ are
related by
\begin{equation}
\langle e^{-2\pi i \bfm\cdot\bfz}\rangle_{p^\prime}  =
 e^{2\pi i \bfm \cdot\bfa} \langle e^{-2\pi i \bfm\cdot\bfz }\rangle_p \,,
\quad
\bfa\in \mathbb{C}^d
\,.
\end{equation}
Therefore, it is sufficient to find positive representations for
$A_\bfn=1$, since other values are generated by a translation (barring
the trivial case $A_\bfn=0$):
\begin{equation}
P_\bfn(\bfx)=1+ e^{2\pi i \bfn\cdot\bfx} 
\,,\quad \bfn\not=\bfzero
\,.
\end{equation}
Let $p_\bfn(\bfz)$, with $\bfz=\bfx+i\bfy$, be a positive
representation of $P_\bfn(\bfx)$. Because $p_\bfn(\bfz)$ is also
periodic with respect to $\bfx$, it can be decomposed in discrete
Fourier modes, $e^{2\pi i\bfk\cdot\bfx}$. The expectation values of
$e^{-2\pi i\bfm\cdot\bfz}$ indicate that the Fourier modes
$\bfk=\bfzero$ and $\bfk=\bfn$ should be present. The mode
$\bfk=-\bfn$ must also be present, for $p_\bfn(\bfz)$ to be real. The
minimum required is thus
\begin{equation}
p_\bfn(\bfz)= h_0(\bfy) + h(\bfy)e^{2\pi i\bfn\cdot\bfx} 
+ h^*(\bfy)e^{-2\pi i\bfn\cdot\bfx},
\quad \bfz=\bfx+i\bfy,
\end{equation}
and $h_0(\bfy)$ real. Furthermore,
\begin{equation}
h_0(\bfy)  \ge 2|h(\bfy)|
\end{equation}
ensures the positivity of $p_\bfn(\bfz)$.

A solution is easily found using the ansatz
\begin{equation}
h(\bfy)=a_1\delta(\bfy-\bfy_1)-a_2\delta(\bfy-\bfy_2)\,,
\quad
a_1,a_2\ge 0 \,.
\end{equation}
At least two delta distributions with weights of opposite sign are
required to satisfy $\langle e^{2\pi i
\bfn\cdot\bfz}\rangle_{p_\bfn}=0$.  Saturation of the bound provides
$h_0(\bfy) = 2|h(\bfy)|$. The remaining conditions $\langle
1\rangle_{p_\bfn}=\langle e^{-2\pi i \bfn\cdot\bfz}\rangle_{p_\bfn}=1$
give the solution
\begin{widetext}
\begin{equation}
p_\bfn(\bfz)=\frac{
2e^{2\pi \bfn\cdot\bfy_1} \cos^2(\pi \bfn\cdot\bfx)\delta(\bfy-\bfy_1)
+
2e^{2\pi \bfn\cdot\bfy_2} \sin^2(\pi \bfn\cdot\bfx)\delta(\bfy-\bfy_2)}
{e^{2\pi \bfn\cdot\bfy_1}+e^{2\pi \bfn\cdot\bfy_2}}\,,
\quad \bfn\not=\bfzero \,,
\end{equation}
\end{widetext}
where $\bfy_1, \bfy_2 \in \mathbb{R}^d$ are only constrained to satisfy
\begin{equation}
2=e^{2\pi \bfn\cdot\bfy_1}-e^{2\pi \bfn\cdot\bfy_2} \,.
\end{equation}
The support of $p_\bfn(\bfz)$ is $\mathbb{R}^d +\mathbb{R}^d$ (or
rather $T^d + T^d$).  This is quite localized as compared to
$\mathbb{C}^d$ or $\tilde{T}^d$.  The localization within this family
of PDFs increases (meaning lower Shannon entropy relative to $T^d +
T^d$\,\footnote{The Shannon entropy of a probability density function
$P$ relative to another PDF $P_0$ is defined as $\langle
-\log(P/P_0)\rangle_P$.  The larger the entropy the less localized is
the probability.  }) by taking $\bfn\cdot\bfy_2\to -\infty$ so that
the branch $\bfy=\bfy_1$ dominates the density function.  The limit
does not exist, even in the weak sense, since the test functions
$e^{-i\bfm\cdot\bfz}$ are of rapid growth on $\mathbb{C}^d$ away from
the real axis.

\section{Concluding remarks}

We have shown that very general complex PDFs can be represented by
positive representations upon analytical extension of the original
manifold.  This allows to sample them as required in the Monte Carlo
method.  The existence of these positive representations was not
granted given the repeated failure of algorithms such as the complex
Langevin approach when applied to general complex weight functions.  A
virtue of our construction is that it does not depend on the
analytical extension of $P(\bfx)$ itself.  In general one cannot expect
$P(\bfz)$ to be well behaved on the complex manifold.  This problem
affects severely the complex Langevin approach.  Nevertheless, it
should be clear that our construction, as it stands, is more of formal
interest than of practical use.  First, the normalization of $P(\bfx)$ is
generally not known; the normalization is needed to choose $P_0(\bfx)$
and so to obtain the difference $\rho(\bfx)$.  Second, given $\rho(\bfx)$,
$F(\bfx)$ is not easy to construct.  In fact, in practical cases, such as
large lattices, only algorithms that are local (i.e., not much more
than nearest neighbors) have a chance to be viable. This does not seem
to be the case of, say, (\ref{eq:5.6}).

Another limitation is related to importance sampling which was one of
the main problems appearing in the reweighting method, (\ref{eq:1.1}).
Strictly speaking this problem does not exist if a pure Monte Carlo
method is applied to $p(\bfz)$.  However, a lack of importance sampling
manifests itself as an enhancement in the fluctuations of averages
over samples.  Because the positive representation $p(\bfz)$ is not
unique there are different representations of a given $P(\bfx)$, all of
them with the same expectation values on analytic functions (but of
course with different expectation values for arbitrary, non analytic,
test functions).  For instance, the convolution of a given
representation $p(\bfz)$ with a positive function $C(\bfz)$ with radial
symmetry and rapidly decreasing at infinity yields a new
representation $p^\prime(\bfz)$ (this is because analytic functions are
invariant under such convolution) \cite{Salcedo:1996sa}.  The new
representation will be wider, i.e., less localized, than the original
one.  As a consequence, although the expectation values of (analytic)
observables will be equal, $\langle f\rangle_p = \langle
f\rangle_{p^\prime}$, the dispersion will be different, being larger
for $p^\prime(\bfz)$.  Therefore, it becomes crucial in the
representation approach to find representations as localized, with
entropy as small, as possible.

This can be seen in another way. Eq. (\ref{eq:5.2}) indicates that the
Monte Carlo method can be applied in the following manner.  First, a
sample of $\bfx$ is generated from $P_0(\bfx)$, and then, for each
$\bfx$ the field $\bfH(\bfx)$ is computed and a sample of $z_1$ is
generated from $q_1(z_1)$ to compute $\langle
f(\bfx+z_1\bfH(\bfx))\rangle_{q_1}$.  This $\bfx$-dependent average is
then itself averaged over the sample of $P_0(\bfx)$ to finally yield
$\langle f(\bfz)\rangle_p$.  However, in most, if not all, cases it
will be more sensible to compute the average over $q_1$  exactly,
\begin{equation}
\langle f(\bfx+z_1\bfH(\bfx))\rangle_{q_1} = 
f(\bfx)-\bfH(\bfx)\cdot\bfnabla f(\bfx) \,,
\end{equation}
rather than using Monte Carlo.  So a better approach
is to directly compute
\begin{equation}
\langle f\rangle_P = \langle f-\bfH\cdot\bfnabla f\rangle_{P_0} \,,
\label{eq:7.1}
\end{equation}
(Interestingly, this formula does not require the analytical extension
of $f(\bfx)$.) Such an approach is  to be compared with the standard one,
(\ref{eq:1.1}),
\begin{equation}
\langle f\rangle_P = \langle f+P_0^{-1}(P-P_0) f\rangle_{P_0}
\label{eq:7.2}
\end{equation}
where, as in (\ref{eq:7.1}), we have assumed $P_0$ to be normalized
(not the case in practice).  The two constructions are rather similar
(formally $\bfH=P_0^{-1}\bfnabla^{-1}(P-P_0)$) so a priori there is no
compelling reason to expect that the importance sampling problem gets
better in (\ref{eq:7.1}) than in (\ref{eq:7.2}). [An exception would
be perhaps the cases in which $f$ is particularly flat, since then
this good property is enhanced by the derivative in (\ref{eq:7.1})].

Despite these critical remarks, it remains the fact that, as shown
here, the representation problem admits a solution for very general
complex PDFs $P(\bfx)$.  Clearly, representability is a sine qua non
condition for the success of any other approach based on analytical
extension, whatever the method used in the construction of the
positive representation.

From the mathematical point of view, several interesting problems pose
themselves.  One is finding more general construction methods, in
addition to the one presented here based on the charge-dipole
decomposition.  Another is carrying out the construction for more
general manifolds and more general complex PDFs.  Also challenging is
the problem of finding positive representations of minimal entropy for
a given $P(\bfx)$: as mentioned before, by convolution it is always
possible to increase the entropy, but there is no general mechanism to
decrease it, so a minimum value is to be expected. [A similar entropy
minimization problem has been found in a different context in
\cite{Salcedo:1998pe}.] If the positivity condition on $p(\bfz)$ is
relaxed, a quite localized representation is
$p(\bfz)=P(\bfx)\delta(\bfy)$. Quite likely, this is the optimum solution if
$P(\bfx)$ is positive.  In the general case of complex $P(\bfx)$, imposing
$p(\bfz)$ to be positive will probably imply a greater delocalization on
the complex manifold.  Finally, one can try to extend the
representativity problem.  In fact in our discussion we have made use
of a map $K$ from the space of test functions $f(\bfx)$ to that of
functions ${\tilde f}(\bfz)$ by means of analytical extension, as well as
the adjoint map $K^\dagger$ which is a projection from the set of
PDFs $p(\bfz)$ on the complex manifold to that of complex
density $P(\bfx)$: $K|f\rangle=|\tilde f\rangle$, $\langle K^\dagger
p|=\langle P|$ (note that the map $p\mapsto P$ is single-valued
although $P\mapsto p$ is not). That is
\begin{equation}
\langle f\rangle_P=
\langle P|f\rangle=
\langle K^\dagger p|f\rangle
=
\langle p| K f\rangle
=
\langle p| \tilde{f}\rangle
=
\langle \tilde{f}\rangle_p
\end{equation}
From this  point of  view, more general  representations $K$  could be
sought which could be of practical interest.

\begin{acknowledgments}
This work is supported in part by funds provided by the Spanish DGI
and FEDER funds with grant FIS2005-00810, Junta de Andaluc{\'\i}a
grants FQM225-05, FQM481 and P06-FQM-01735 and EU Integrated
Infrastructure Initiative Hadron Physics Project contract
RII3-CT-2004-506078.
\end{acknowledgments}

\appendix

\section{Polynomial growth of $\bfH(\bfx)$}
\label{app:A}

Here we want to show that $P_0(\bfx)$ can be chosen in such a way that
$\bfH(\bfx)$ is bounded by a polynomial.  We take it as evident that
for any complex Schwartz function $P^\prime$ on $\mathbb{R}^d$ there
is a positive function $P_0^\prime$ such that (i) $P_0^\prime$ is in
Schwartz space, (ii) $P_0^\prime(\bfx)\ge|P^\prime(\bfx)|$ for all
$\bfx\in\mathbb{R}^d$, and (iii) $P_0^\prime$ is a decreasing
function, i.e., $P_0^\prime(\bfx_1) \le P_0^\prime(\bfx_2)$ if
$|\bfx_1|\ge |\bfx_2|$.\footnote{To obtain $P_0^\prime$ consider the
auxiliary function
$$
P_0^{\prime\prime}(\bfx)
= \sup\{|P^\prime(\bfy)|, |\bfy|\ge|\bfx|\}\,.
$$ By construction $P_0^{\prime\prime}$ satisfies all conditions (i-iii),
except perhaps the property of being smooth. It seems obvious that
$P_0^{\prime\prime}$ can be suitable smoothed out to obtain a valid
$P_0^\prime$.} We apply this property to the function
$P^\prime(\bfx):=|\bfx|^{2d}P(\bfx)$ and take
$P_0(\bfx):=|\bfx|^{-2d}P_0^\prime(\bfx)$ for $|\bfx|\ge R$, for a
sufficiently large $R$.  The definition of $P_0(\bfx)$ is completed in
the region $|\bfx|<R$ so that it is smooth and normalized. Therefore,
for $|\bfx|\ge R$, $|P(\bfx)|\le P_0(\bfx)$ and
\begin{eqnarray}
&& |\rho(\bfx)| \le 2 P_0(\bfx)\,,
\nonumber\\
&& P_0(\bfy) \le \frac{|\bfx|^{2d}}{|\bfy|^{2d}}P_0(\bfx) \,,
\quad |\bfy| \ge |\bfx| \,. 
\label{eq:A.1}
\end{eqnarray}
(As in Section \ref{sec:5}, $\rho=P-P_0$.)  In order to use these
inequalities we recall that in (\ref{eq:5.6}) the support of the
positive function $C(\bfa)$ is compact, so $\bfa$ lies inside a ball
of some size $R_C$, and moreover $\lambda\ge 1$. Thus the
inequalities (\ref{eq:A.1}) apply at the point
$\lambda(\bfx-\bfa)+\bfa$ provided $R>R_C$. This allows to write
\begin{eqnarray}
 |\bfF(\bfx)| &\le& \int_1^\infty d\lambda \,\lambda^{d-1}
\int_{\mathbb{R}^d} d^da\, C(\bfa)
\nonumber\\
&&\times
\frac{2|\bfx-\bfa||\bfx|^{2d}}{|\lambda(\bfx-\bfa)+\bfa|^{2d}}P_0(\bfx) 
\,,
\quad 
|\bfx|\ge R \,.
\end{eqnarray}
On the other hand $|\lambda(\bfx-\bfa)+\bfa| 
\ge \lambda(|\bfx|-|\bfa|)$, hence
\begin{eqnarray}
 |\bfF(\bfx)| &\le& 2\int_1^\infty d\lambda \,\lambda^{-d-1}
\int_{\mathbb{R}^d} d^da\, C(\bfa)
\nonumber\\
&&\times
\frac{|\bfx-\bfa||\bfx|^{2d}}{(|\bfx|-|\bfa|)^{2d}}P_0(\bfx) 
\nonumber\\
&\le&
\frac{2}{d}\frac{(|\bfx|+R_C)|\bfx|^{2d}}{(|\bfx|-R_C)^{2d}}P_0(\bfx) 
\,,
\quad 
|\bfx|\ge R \,.
\end{eqnarray}
\vbox{
As a consequence,
\begin{eqnarray}
|\bfF(\bfx)| &\le&  k|\bfx|P_0(\bfx)
\,, 
\quad \text{for~~} |\bfx|\ge R \,,
\nonumber \\
k&=&\frac{2}{d}\frac{1+R_C/R}{(1-R_C/R)^{2d}}
 \,,
\end{eqnarray}
and $\bfH(\bfx)$ is bounded by $k|\bfx|$.
}

%\vfill
%\eject

%\bibliography{Refs}
%\bibliographystyle{h-physrev4}

\end{document}